# The Responsibility Quantification (ResQu) Model of Human Interaction with Automation

Nir Douer and Joachim Meyer, *Senior Member, IEEE*

*Abstract* — Intelligent systems and advanced automation are involved in information collection and evaluation, in decision-making and in the implementation of chosen actions. In such systems, human responsibility becomes equivocal. Understanding human causal responsibility is particularly important when systems can harm people, as with autonomous vehicles or, most notably, with autonomous weapon systems (AWS). Using Information Theory, we developed a responsibility quantification (ResQu) model of human causal responsibility in intelligent systems and demonstrated its applications on decisions regarding AWS. The analysis reveals that human comparative responsibility for outcomes is often low, even when major functions are allocated to the human. Thus, broadly stated policies of keeping humans in the loop and having meaningful human control are misleading and cannot truly direct decisions on how to involve humans in advanced automation. The current model assumes stationarity, full knowledge regarding the characteristic of the human and automation and ignores temporal aspects. It is an initial step towards the development of a comprehensive responsibility model that will make it possible to quantify human causal responsibility. The model can serve as an additional tool in the analysis of system design alternatives and policy decisions regarding human causal responsibility, providing a novel, quantitative perspective on these matters.

*Note to Practitioners* — We developed a theoretical model and a quantitative measure for computing the comparative human causal responsibility in the interaction with intelligent systems and advanced automation. Our responsibility measure can be applied by practitioners (system designers, regulators, etc.) to estimate user responsibility in specific system configurations. This can serve as an additional tool in the comparison between alternative system designs or deployment policies, by relating different automation design options to their predicted effect on the users' responsibility. To apply the model (which is based on entropy and mutual information) to real-world systems, one must deduce the underlying distributions, either from known system properties or from empirical observations, taken over time. The initial version of the model we present here assumes that the combined human-automation system is stationary and ergodic. Real-world systems may not be stationary and ergodic or cannot be observed sufficiently to allow accurate estimates of the required input of multivariate probabilities, in which case the computed responsibility values should be treated with caution. Nevertheless, the construction of a ResQu information flow model, combined with sensitivity analyses of how changes in the input probabilities and assumptions affect the responsibility measure, will often reveal important qualitative properties and supply valuable insights regarding the general level of meaningful human involvement and comparative responsibility in a system. *Index Terms*— Analytical models, Artificial intelligence, autonomous systems, decision making, human–computer interaction (HCI), information theory, Intelligent systems, responsibility.

Manuscript first submitted October 2018; revised and resubmitted April 2019; conditionally accepted and revised August 2019; conditionally accepted and revised December 2019; Accepted January 2020. This work was partly supported by the Israel Science Foundation Grant 2029/19. *(Corresponding author: Joachim Meyer).*

N. Douer and J. Meyer are with the Department of Industrial Engineering at Tel Aviv University, Ramat Aviv, Tel Aviv 69978, Israel (e-mails: nirdouer@mail.tau.ac.il, jmeyer@tau.ac.il).

## I. Introduction

Advanced automation and intelligent systems have become ubiquitous and are major parts of our life. Financial markets largely function through algorithmic trading mechanisms [1, 2], semiconductor manufacturing is almost entirely automated [3], and decision support systems and aids for diagnostic interpretation have become part of medical practice [4, 5]. Similarly, in aviation, flight management systems control almost all parts of the flight [6, 7], and in surface transportation, public transportation is increasingly automated, and the first autonomous cars appear on public roads [8, 9]. In these systems, computers and human share the execution of different functions, such as the collection and evaluation of information, decision-making and action implementation.

As these intelligent systems become more advanced, the human comparative responsibility for outcomes becomes equivocal. For instance, what is a human's responsibility when all information about an event arrives through a system that collects and analyzes data from multiple sources, without the human having access to any independent sources of information? If the human receives an indication that a certain action is needed, and accordingly performs the action, should the human be held responsible for the outcome of the action, if it causes harm?

Human responsibility is particularly important when system actions can possibly injure people, as may be the case with autonomous vehicles. It becomes crucial when such harm is certain, as with autonomous weapon systems, deliberately designed to inflict lethal force.

So far, the subject of human responsibility was investigated from philosophical, ethical, moral and legal perspectives. However, we still lack a quantitative engineering model of human responsibility. To address this need, we developed the Responsibility Quantification (ResQu) model that enables us to compute human responsibility in the interaction with intelligent systems and automation. We will demonstrate its application on the example of autonomous weapon systems, because this issue raises particular public concerns. However, the model is applicable wherever intelligent systems and automation play a major role.

### A. Responsibility in human-automation interaction

Philosophical and legal research has dealt extensively with the concept of responsibility, investigating its different facets, namely role responsibility, causal responsibility, liability (or legal responsibility) and moral responsibility [10-12]. When discussing human interaction with intelligent systems and automation, *role responsibility* relates to assigning specific roles and duties to the operator, for which the operator is accountable. However, this role assignment does not specify the causal relations between the operator's actions and possible consequences and outcomes. This relation is better defined by *causal responsibility*, which describes the actual human contribution to system outcomes.

A large literature in psychology, such as attribution theory, sees causal responsibility as an essential primary condition for the attribution of blame and praise [13-17]. Causal responsibility is also a major factor in the way legal doctrines determine liability, punishments and civil remedies in criminal and tort law [18-20].

So far, causal responsibility was usually associated with people - a person or an organization was seen as more or less responsible for a particular event. When an event involved technology, the responsibility was usually with the user, unless some unforeseeable circumstances caused some unexpected outcome. Manufacturers of systems could also be held responsible if, for instance, they failed to install proper safeguards.

The field changed with the introduction of automation, defined as the system performing parts, or all, of a task that was, or could have been, performed by humans [21]. The ability to control a system and the resulting consequences is a necessary condition for assigning responsibility. However, humans may no longer be able to control intelligent systems and advanced automation sufficiently to be rightly considered responsible. As the level of automation and system intelligence increase, there is a shift towards *shared control*, in which the human and computerized systems jointly make decisions or control actions. These are combined to generate a final control action or decision. There may also be *supervisory control*, in which the human sets high-level goals, monitors the system and only intervenes if necessary [22]. In coactive designs, humans and systems engage in joint activities, based on supporting interdependence and complementary relations in performing sensing, planning, and acting functions [23, 24]. Moreover, in advanced systems, which incorporate artificial intelligence, neural networks, and machine-learning, developers and users may be unable to fully control or predict all possible behaviors and outcomes, since their internal structure can be opaque (a "black box") and sometimes can yield odd and counterintuitive results [25, 26].

Consequently, humans' causal responsibility in intelligent or highly automated systems becomes equivocal and cannot be separated from the causal contribution of the system's configuration and reliability. The automated system itself (or its developers) may be perceived as sharing some of the responsibility [27, 28]. This understanding resembles the legal concept of comparative responsibility, a doctrine of tort law that divides fault among different parties [29-31].

The limited human ability to influence the outcomes when interacting with highly automated systems may create a discrepancy between role responsibility (i.e. the duties of the human operator, that he or she are accountable for), and causal responsibility, which describes the actual influence of the human actions on system outcomes. There are several possible causes for this discrepancy. First, humans may lack the *authority* to take the actions necessary to fulfill their role, and consequently, they may have limited ability to influence the system outcomes. This is known as *responsibility-authority double binds,* which describes a situation in which the human operator is assigned a specific role, but is not granted sufficient authority to act and control the processes that lead to the outcomes [32, 33]. Secondly, as we have described, intelligent systems and advanced automation may limit the human ability to control and take the necessary actions to influence the outcomes, even when the human is granted the authority to act (e.g. when they include opaque processes and interfaces). Lastly, since causal responsibility is measured considering the outcomes, it is influenced by uncertainties and probabilistic aspects that are not part of authority (which is defined and granted beforehand). For example, the human may have sufficient authority to act or override any system decision, but due to probabilistic factors, related to the automation or the environment, the human's actions may have, in fact, only minor impact on the probability distribution of the outcomes.

In either case, the human may be considered fully legally responsible for adverse outcomes, even when not having sufficient control to prevent them or when contributing very little to create these outcomes. Hence, a measure of the marginal causal human responsibility can provide a more adequate description of the human's contribution to the outcomes.

To conclude, the rapid developments in technology create an inevitable *responsibility gap* in the ability to divide causal responsibility between human and advanced automated or intelligent systems. This gap cannot be bridged using traditional concepts of responsibility [34-36].

One attempt to quantify causal responsibility, in multiple-agent contexts, is a recent structural model (often referred to as the "counterfactual pivotality model") [37]. The model considers a group of agents, a given set of their selected actions and the resultant combined outcome. An agent is considered more responsible for the outcome if the agent's action was pivotal (made a difference) for creating the outcome, considering all other agents' actions, and also by whether it would have made a difference in other possible (counterfactual) situations. For a given set of action selections and an outcome, the model defines an agent's causal responsibility as $1/(N+1)$ where $N$ denotes the minimal number of changes that have to be made to the original situation (mainly changes in other agents' actions), in order to make the agent's current action selection pivotal. The model has been used in a number of cognitive physiology studies [38-41], but it has limited value for quantifying human involvement in intelligent systems, because it is deterministic and quantifies the *retrospective* causal responsibility for a specific known set of actions and an outcome. Conversely, in order to assist system design, the focus should be on *prospective* causal responsibility, which is the average causal contribution of the human over distributions of future events and complex human-machine interactions, in a probabilistic world. In addition, most human and machines do not act as full substitutes or full complements, but rather perform joint, interdependent functions. These factors preclude the use of the model's responsibility measure for dividing causal responsibility between human and intelligent systems.

In this paper we aim to address the above difficulties and gaps, by developing a responsibility quantification (ResQu) model of human causal responsibility in intelligent systems. We present a new method to quantify the comparative human causal responsibility. To do so, the ResQu model considers major factors that influence the human ability to control and determine the outcomes, such as authority, system design, human capabilities and environmental factors. The model takes into account probabilistic aspects, by using information theory to analyze the interdependencies within the human-machine system and the environment.

### B. Causal responsibility in autonomous weapon systems

Human causal responsibility is a major issue in the discussions of autonomous weapon systems (AWS). While there are several definitions, the term "autonomous systems" refers to systems which acquire information from the unstructured probabilistic world around them, analyze the information, make decisions and implement them, with limited or no human supervision and control. The processes and outcomes of autonomous systems are probabilistic in nature and may not be predictable. In contrast, automated systems sense and respond deterministically to unambiguous events, using clear repeatable rules [42, 43].

AWS are not necessarily either fully automated or fully autonomous, but autonomous in some of their functions (e.g., drones, missiles and smart munitions with autonomous navigation, surveillance, and terminal guidance functions). However, the emerging autonomous abilities to detect, select and engage targets independently are at the center of the current debate. Such autonomous abilities of critical engagement functions are already implemented in missile defense systems, vehicle "active-protection", sensor-fused and loitering munitions, and are under development for various future offensive weapons [44, 45].

The rapid technological developments in AWS have raised concerns that with increasingly intelligent and autonomous military technologies, humans will become less and less involved in their use. They will be considered or may feel less responsible for lethal outcomes [35, 46-49], opening an unacceptable responsibility gap in AWS [50].

These systems also raise critically important issues of controllability and safety, since in the event of a failure, they could lead to catastrophes, such as mass fratricide or civilian casualties, with limited (or no) human ability to intervene and prevent the adverse consequences [26]. These concerns prompted extensive philosophical, ethical, and legal debates, which elicited calls to restrict and regulate the development of advanced AWS, or even ban their use altogether [50-60].

Governments respond to these worries with the assurance and demands that a human will be kept in the loop, whenever advanced automated systems exert lethal force [46]. The explicit policy of the U.S. Department of Defense is that "Autonomous and semi-autonomous weapon systems shall be designed to allow commanders and operators to exercise appropriate levels of human judgment over the use of force" [61]. In addition, under U.S. policy, supervised autonomous weapon systems may select and engage targets only in local defensive operations, such as protecting land-bases and ships, and fully autonomous weapon systems are limited to application of non-lethal, non-kinetic force [61]. The UK policy is that the operation of weapon systems will always be under human control, and that no offensive systems should be able to prosecute targets without involving a human [62].

These policies are mainly based on philosophical and legal perspectives. Currently, there is an important need for the involvement of different scientific disciplines in the discussions, to support better framing of the debate and to design meaningful policies and regulations [43].

In this paper we demonstrate how the ResQu model can generate new perspectives on current policies by analyzing a basic scheme of an AWS, which automatically detects, classifies and engages targets, but requires some level of human involvement in the engagement process. Despite complying with current U.S. and UK policies, such an AWS is controversial and raises major concerns regarding human responsibility.

### C. Meaningful human control

The United Nations Institute for Disarmament Research and other organizations promoted the need for *meaningful human control* of AWS [63]. This approach aims to ensure that commanders and operators will have enough information to make conscious decisions and can intervene if necessary, and that AWS should be designed to facilitate such meaningful control [64].

It is important to note that the demand to involve humans in automated processes and to facilitate meaningful human control is not unique to AWS [65]. It also applies to other intelligent systems, such as computers, autonomous cars, surgical robotics, and more [66-68].

However, simply putting a human into the loop does not assure that the human will have a meaningful role in the process. There may be cases when the human cannot knowledgably supervise the system, or when the human has to make decisions, based exclusively on input from automated functions that one cannot evaluate independently [69]. System designers often keep humans in the loop to cope with unexpected events, even when the human may be unable to cope with such events. In this case, humans may function as "moral crumple zones", being the ones to carry moral and legal responsibility when the system fails [70, 71].

Currently, there are different, and sometimes contradicting interpretations and policies regarding meaningful human involvement. System designers lack models and metrics needed for systematically addressing the issue of meaningful human control in autonomous systems [43, 72].

The ResQu model can address these needs, by providing an estimate for how meaningful the human involvement in a system is, based on the premise that meaningful human control requires the human to have some causal responsibility for the outcomes.

II. THE RESPONSIBILITY QUANTIFICATION MODEL (RESQU)

*A. A general model of information flow in a human-automation system*

According to Parasuraman et al. [21], a combined human-automation system performs a sequence of four consecutive information processing functions: the acquisition of information, the analysis of that information, the decision what action to take, based on the information, and the implementation of the action. Each of the four functions can be automated, from the lowest level of fully manual performance to the highest level of fully automatic performance.

A model developed by Conant [73] uses n-dimensional Information Theory to analyze the information flow in real-world systems, composed of interacting parts and subsystems. The system acquires input from its environment, and it generates output to the environment. Each variable in the system is a message source, which sends information about its values to other variables. Thereby, the functioning of the system, which is usually formulated as a process of causes, effects, and activities, becomes a network of transmitters, channels, and receivers. With this representation, one can quantify the information flow, causal relations, and statistical dependence between variables and subsystems in terms of Entropy and Mutual Information (also called Transmission).

We integrated Parasuraman et al.'s and Conant's models and created a general model of information flow in a combined human-automation system. Similar to principles of coactive design [23, 24], the integrated model includes both the human and the machine as equal components of the integrated system and supports interdependent human-machine relations in performing sensing, planning, and acting functions. However, differently from coactive design, our model uses information theory to analyze the interactions and interdependencies within the human-machine system and with the environment.

The integrated information flow model will serve us to quantify human responsibility as a function of the system design, selected automation levels, and function allocation.

*B. Notation*

*The System*: Assume a system that consists of two subsystems: an automated module and a human user. Although the terms "system" and "automation" usually carry similar connotations, in the present study the term system refers to the overall system, containing both the human and the automated module subsystems, and it indicates their combined performance. The terms "human" and "automated module" refer to the subsystems and to their specific performance and parameters.

*Environment states*: The system operates in an environment that can be in one of N possible states ($N \geq 2$). Each of the $N$ states can be characterized by $m$ different observable and measurable parameters $E_i$ $(i=1...m)$. Different states have different, but partially overlapping, distributions on each of the values of $E_i$ $(i=1...m)$. Thus, when observing a specific realization of $E_i$ $(i=1...m)$, the current environment state remains uncertain.

*Information acquisition*: The first stage deals with the acquisition and registration of multiple sources of information. Let $Y_i$ and $X_i$ $(i=1...m)$ denote, respectively, the acquired values of $E_i$ $(i=1...m)$ by the automated module and the human. Due to measurement and accuracy limitations of the sensors, the measurements $Y_i$ and $X_i$ $(i=1...m)$ may add uncertainty (or internal noise) to the actual observed value $E_i$, $(i=1...m)$. In addition, not all of the $m$ state-characteristic variables are observable by the human or the automated module. When a certain state characteristic variable $E_i$ is not observable by the human, $X_i$ will not contain any information about $E_i$ (and the same for $Y_i$, when $E_i$ is not observable by the automated module).

*Information analysis*: The second stage deals with manipulation of the acquired information to infer the current environment state. Let $Y_a$ and $X_a$, denote, respectively, $N$ dimension vectors, generated by the automated module and human subsystems, that assign posterior probabilities to each of the possible $N$ environment states, based on the acquired information by each subsystem $Y_i$, and $X_i$ $(i=1...m)$. Depending on the system's automation level and function allocation, the results of one subsystem's information analysis or action selection may serve as an additional source of information for the other subsystem's information analysis.

*Action selection*: In the third stage, decisions are made and actions are selected, based on the results of the information analysis. Let $Y_s$ and $X_s$ denote, respectively, variables of the automated module and the human that correspond to the selection of a preferred action amongst a set of finite action alternatives. For the automated module, $Y_s$ is uniquely defined by the automated module's algorithm, once all input variables $Y_i$, $(i=1...m)$ are acquired, and the analysis algorithm $Y_a$ is executed. For the human, $X_s$ is based on the results of the human information analysis $X_a$, and it depends on characteristics of the human utility function, which relates costs and benefits to different outcomes that may be generated by each action alternative.

*Action implementation*: The fourth and final stage involves the implementation of the selected action. Let $Z$ denote the implemented action. We assume that the implemented action $Z$ only depends on the actions the automation and the human selected, $Y_s$ and $X_s$, and on the relative amount of human versus automatic impact on generating a response. Dictated by the system configuration and the automation level, Z may be entirely determined by the human action selection, by the automated module, or by a combination of the two. In systems that incorporate adaptive automation (dynamic function allocation), the determination of Z may change, depending on the identified environment state. This may be the case, for example, in systems where automation can override human actions when it identifies a critical emergency that is beyond human response capabilities, such as automatic emergency breaking systems in cars.

Fig. 1 presents a schematic depiction of system variables and information flows, between the human and the automation and across the different information processing functions.

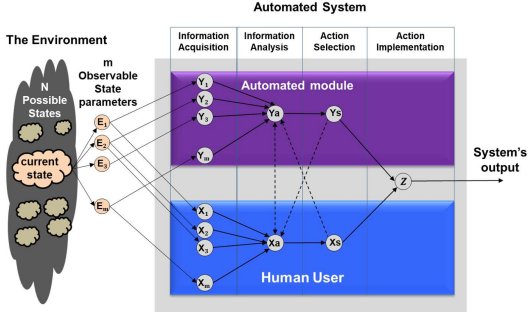

Fig. 1. General model of information flow in an automated human-machine system. Dashed lines represent possible information transfer between the human user and the automated module

The general information flow model, presented above, portrays possible system variables and information flows in a human-automation system. However, depending on the system architecture, in actual systems some of the variables and information flow routes may not exist. For example, alerting systems that indicate a potential hazard or identify other pre-specified conditions (e.g. in industrial control rooms, medical equipment, anti-malware detection etc.) [7, 78-80], and AI applications, which perform complex categorization and classification tasks (e.g. consumer segmentation, automated recommendations, targeted advertising, etc.), conduct only the analysis function and present their results to the human for further decision making and action selection. In these systems, the automated system itself may recommend an action, but the final action selection is left to the human. Fig. 2 presents an example of the information flow in such systems.

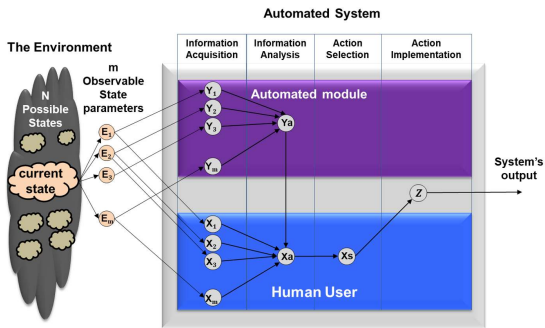

Fig. 2. An example of information flow in human interaction with recommendation systems, such as alert or classification systems.

The general ResQu model can be similarly applied to analyze the information flow in other systems with various levels of automation and types of human control, such as shared control or supervisory control

*C. Defining a responsibility measure*

We measure human responsibility by quantifying the unique comparative share of the human in determining the distribution of the system output Z (the implemented action). We do so by computing the proportion of the output distribution that does not result from automation, and thus represents the unique share of human contribution in determining the system's output. Using Information Theory [74, 75] we define the comparative causal human responsibility for the system output Z as

$$Resp(Z) \stackrel{\text{def}}{=} \frac{H(Z/Y_1...Y_m,Y_a,Y_s)}{H(Z)} \quad (1)$$

where *H(X)* is Shannon's entropy, which is a measure of uncertainty related to a discrete random variable *X*, defined as:

$$H(X) \stackrel{\text{def}}{=} -\sum_{x\in\chi} p(x)log_2 p(x) \quad (2)$$

and *H(X/Y)* is the conditional entropy, which is a measure of the uncertainty remaining about a variable *X* conditioned on another random variable *Y*:

$$H(X/Y) \stackrel{\text{def}}{=} -\sum_{y\in\gamma} p(y) \sum_{x\in\chi} p(x/y) \, log_2 p(x/y) \quad (3)$$

The conditional entropy in the numerator of *Resp(Z)*, $H(Z/Y_1 ... Y_m, Y_a, Y_s)$ is the remaining uncertainty about *Z* (the overall system output), conditioned on all automation information processing functions. In our model, there is no internal noise or blockage of information within the combined system, so this remaining uncertainty is only due to human subsystem variables. Thus, the ratio of the conditional entropy and the entropy of *Z* quantifies the unique comparative human contribution to the distribution of the system output *Z*.

By definition, $Resp(Z) \in [0,1]$. $Resp(Z)=1$ iff $H(Z/Y_1 ... Y_m, Y_a, Y_s) = H(Z)$. This occurs if, and only if, the system output variable *Z* is independent from the automation variables. In that case, all uncertainty about *Z* is completely resolved by the human, and thus the human is fully responsible for the system output. $Resp(Z)=0$ iff $H(Z/Y_1 ... Y_m, Y_a, Y_s) = 0$. This happens if, and only if $Y_1 ... Y_m, Y_a, Y_s$ completely determine *Z* without any unique contribution by the human. Values between 0 and 1 represent intermediate levels of unique human contribution to the overall output (i.e. level of meaningful human involvement), given the automation performance. Using Shannon's entropy, *Resp(Z)* averages the comparative human contribution over all possible states in the environment, their distribution on the set of measurable parameters, and the resultant distributions of human and automation parameters and system outputs.

Our responsibility measure is related to Theil's uncertainty coefficient, *U(X/Y)*, which is a measure of the association between two variables *X* and *Y* [76, 77]. Theil's uncertainty coefficient computes the relative reduction in the uncertainty of a variable $X$ due to the knowledge of another variable $Y$:

$$U(X/Y) \stackrel{\text{def}}{=} \frac{I(X:Y)}{H(X)} = \frac{H(X)-H(X/Y)}{H(X)} \quad (4)$$

where $I(X:Y)$ is the mutual information between $X$ and $Y$. Theil's uncertainty coefficient is more general than the notion of statistical correlation, since it can be used to measure complex, not necessarily linear associations, as well as associations between nominal variables. It has values between 0 and 1. It equals 0 iff the variables are statistically independent and share no mutual information, and it equals 1 iff knowledge about the value of $Y$ fully enables one to predict the value of $X$.

Our approach differs from Theil's coefficient in that we do not simply measure the association between two variables. Our ResQu responsibility value measures complex associations in a compound human-automation system, characterized by multiple variables and inter-dependencies. Also, Theil's coefficient focuses on the relative *reduction* in the uncertainty of a variable, given knowledge about another variable, while our ResQu score computes the relative *remaining* uncertainty, given knowledge about other variables.

It is important to note that *Resp(Z)* is based on information theory concepts of entropy and mutual information, which were developed under the mathematical constraints of stationarity and ergodicity. Thus, the combined human-automation system must also be assumed stationary and ergodic. The practical implications of these assumptions are discussed in the next section.

### III. AN APPLICATION OF THE RESQU MODEL

#### A. General

The ResQu model, presented in the previous section, is an abstract model of information flow in a combined human-automation system. Details of how the human and the automation communicate, decide on actions, and resolve discrepancies are abstracted through different variables that characterize the environment, the human and the automation activities during the execution of four information processing functions. The model quantifies the comparative human responsibility by analyzing interdependencies between the various system variables and by determining the human relative contribution to the distribution of combined system outputs.

The ResQu model is not only an abstract theoretical model. It can also be used to analyze responsibility in real-world interactions with automation and intelligent systems. To calculate the responsibility measure for real-world systems, one must infer the underlying distributions of system variables from known properties or from empirical observations, collected over time. As we have stated, the initial version of the model we present here assumes that the combined human-automation system is stationary and ergodic. When applying the model, one needs to keep in mind that real-world systems may not be stationary and ergodic or cannot be observed sufficiently to allow accurate estimates of the multivariate probabilities [73]. Nevertheless, we believe that for these systems, the construction of a detailed ResQu information flow model, combined with sensitivity analyses of how changes in the input probabilities and assumptions affect the responsibility measure, will often provide useful insights regarding the comparative human contribution to the outcomes.

#### B. Information flow model for AWS

We present an application of the ResQu model to a large family of decision support systems (DSS), which automatically classify input to one of two or more categories and may also recommend an action. We demonstrate the use of the model on the schematic example of an AWS, which automatically detects, classifies and engages targets, but alerts the operator and requires some level of human involvement during the engagement process, either in the loop or on the loop.

We model the human control of the AWS in a simplified manner, which does not capture all nuances of function allocation and human control [24, 26, 69, 72, 81, 82]. We also don't explicitly consider uncertainties, related to human factors, that may constrain the human performance, such as variables from the Opportunity-Willingness-Capability (OWC) model (e.g., load, stamina, stress, knowledge, experience or training, cognitive or physical skills, emotional state, etc.) [83-86].

To calculate the human responsibility, we need to make assumptions regarding the probabilistic distributions and the interdependencies of the different variables that characterize the environment, the human, and the AWS. One way to do so is to use the assumptions and formulation of Signal Detection Theory (SDT) [87-89]. This is a well-established approach to measure the ability to differentiate between information-bearing signals (or stimuli) and random noise and to decide on a proper response. SDT has applications in many fields, such as psychology, decision-making, telecommunications, medical diagnostics, biology, alarm management, machine learning (statistical classification), and military (e.g. in radar research).

We used an equal variance Gaussian SDT model to represent the probabilistic nature of the representative AWS, the environment in which it operates, the human activities and interactions with the automation and the incentives, which influence response selection, in a manner described below (see the Appendix for a detailed description). We present the combined human-automation system as a sequence of consecutive information processing functions (information acquisition, analysis to infer the current environmental state, selection of actions, and implementation).

*Environment states*: Assume that an AWS operates in an environment with only two types of entities: targets, which should be engaged, and noise, which are entities that resemble targets, but which should not be engaged. Engagement of noise entities leads to undesired costs, such as collateral damage, fratricide or the waste of expensive or limited ammunition on false targets. The relative frequency of targets in the environment is $P_t$, and the relative frequency of noise is $1-P_t$.

The overall operational goal of the AWS is to detect, identify and engage targets, while avoiding the engagement of noise. To do so, the system automatically scans specific designated areas

(e.g., specific geographic regions or air sectors) for the presence of entities. We assume that the system detects all entities in its vicinity with certainty. Hence, the main challenge for the AWS is the classification of each detected entity as target or noise. The classification of entities relies on the combined performance of two subsystems - a human operator and an automated module.

*Information acquisition*: Target and noise entities have physical characteristics that human senses or other sensors can discern to some extent (e.g., optical, thermal electromagnetic and acoustic signature, mobility characteristics, etc.). Denote by *e*, the set of the observable physical characteristic of the state of the world. We assume that the human operator and the automatic module each observe a different uncorrelated measurable property of the state of the world, based on *e*. The distributions of these properties for target and noise are Gaussian with equal variance, but different means, allowing some discrimination between the two types of entities. However, the distributions overlap, so classifications of observed entities as target or noise are uncertain.

*Information analysis*: In this stage, both the human and the automated module try to infer the current environmental state. In SDT formulation, a detector's ability to classify observations is its *detection sensitivity*. For Gaussian, normal distributions, the human and the automated module have, respectively, detection sensitivities of $d'_{Human}$ and $d'_{Automation}$, where $d'$ is the distance between the means of the signal and noise distributions, measured in standard deviations.

*Action selection*: In this stage, decisions are made, based on the results of the information analysis. These decisions are susceptible to motivation, costs, strategy, etc. In SDT formulation, the action selection is characterized by the *response criterion* (also called response bias), which defines a detector's tendency to classify events as signal or noise. Each response criterion specifies a threshold value. The detector classifies events as targets when they are above the threshold and as noise otherwise. In actual systems, this threshold value is programmed into the automation algorithms. It also reflects the human bias in decision making, depending on the likelihood of targets and the costs and benefits of different outcomes.

According to SDT, the optimal response criterion maximizes the expected value of a payoffs scheme in which $V_{TP}$, $V_{FP}$, $V_{TN}$, and $V_{FN}$ represent, respectively the values associated with correct target classification (True Positive), incorrect target classification (False Positive) when noise is falsely classified as a target, correct classification of noise as noise (True Negatives), and false classification of a target as noise (False Negatives). The payoff scheme, which represents the values of the outcomes in actual systems, will usually not be in monetary terms. Rather, it expresses some assessment of the relative utility of outcomes in terms of costs and benefits (for instance, associating a very high cost, $V_{FP}$, to cases in which a civilian entity is falsely classified as a legitimate target for engagement). These payoffs can reflect the values human operators or system designers associate with outcomes, but they can also reflect the values the organization that deploys the system associates with outcomes.

It is important to note that, due to possible differences in the preferences, the values system designers associate with different outcomes may or may not be identical to the human operators' values. This may lead to differences in the action selection preferences between the human operator and the automation. Hence, we assume that the human and the automated module each have response criteria ($\beta_{Human}$ and $\beta_{Automation}$), defining their bias when classifying an entity as a target or noise

The automated module performs independent binary classifications with its detection sensitivity and preset response criteria. Let *Y* denote its classification result, either target or noise, which may include correct or incorrect classifications of targets and noise. We assume that the engagement process itself is mostly automatic, but it requires some level of human involvement, whether in the loop or on the loop. In human in the loop control, we assume that whenever the automation classifies an entity as a target, the engagement process will only proceed if the human actively authorizes the engagement. The engagement halts if the human decides to abort and remains passive (does not authorize the engagement). In human on the loop control, we assume that whenever the automation classifies an entity as a target, the engagement proceeds automatically, as long as the human remains passive, and halts only if the human actively aborts it. In addition, in both types of control, the human can always decide to engage an entity, even if the automation classified it as noise. Thus, in both cases the human has to decide whether to engage or to abort. To do so, the human combines the information from the automated module with additional information the human has. The only actual difference is whether an active response is required to implement the chosen action, or whether the human can remain passive.

Let *X* denote the human action selection, either to engage or to abort. According to SDT, when aided by such an automated module, a rational, payoff-maximizing human should use two different response criteria: one is used when the automated module classifies an entity as target and the other when the automated module classifies it as noise. The differential adjustment of the response criteria depends on the human's assessment of the automated module's capabilities. When using a reliable AWS with high capabilities, the human should adopt a lower cutoff point when the system classifies an entity as a target, which would increase the tendency to engage, and a higher cutoff point when the system classifies an entity as noise, which would increase the tendency to abort.

*Action implementation*: This final stage involves the implementation of the action the human selected. If the human chooses to engage an entity, the system conducts the rest of the engagement process automatically (e.g., missile lock on target and missile firing). Let *Z* denote the outcome of the integrated system. This outcome represents whether a detected entity was eventually engaged or not. It is important to note that in both types of human control, humans have the final word and can always override and alter the automated module's recommendation, based on their own information analysis and action selection processes. Thus, in the portrayed system, *Z* is strictly determined by the results of the human action selection process.

To conclude, in our simple representative scheme of human interaction with an AWS, the investigation of the four information processing functions, within the combined human-machine system, can be reduced to analyzing three variables and their inter-dependencies: *Y* (the classification result of the automated module), *X* (the human action selection) and *Z* (the outcomes). Fig. 3 depicts the information flow, into, within, and out of the combined human-machine system.

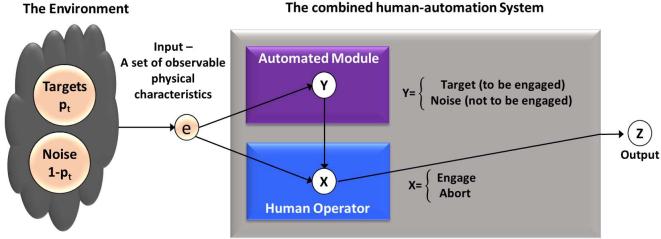

Fig. 3. Information flow and parameters of an AWS system which detects classifies and engages targets automatically but requires human involvement, before or during the engagement process

### C. Defining responsibility measures for the AWS

The information flow and system structure, shown in Fig.3, enable us to simplify the general formula for *Resp(Z)* in (1), to:

$$Resp(Z) \stackrel{\text{def}}{=} \frac{H(Z/Y)}{H(Z)} \quad (5)$$

The conditional entropy in the numerator is the remaining uncertainty about *Z* (whether a detected entity was engaged or not), conditioned on the result of the automation classification. This remaining uncertainty is due to the human actions. Hence, the ratio in (5) quantifies the unique comparative human contribution in determining the distribution of the system engagement output variable *Z*. *Z* is directly determined by the human action selection *X* (i.e. whether the human chose to engage an entity or not), which in turn, is influenced by the information from the automation, *Y* (see Fig. 3).

Hence, we can rewrite (5) as:

$$Resp(Z) \stackrel{\text{def}}{=} \frac{H(Z/Y)}{H(Z)} = \frac{H(X/Y)}{H(X)} \quad (6)$$

The conditional entropy in (6) can be written explicitly as:

$$Resp(Z) = \frac{H(X/Y)}{H(X)} = \frac{H(X,Y) - H(Y)}{H(X)} \quad (7)$$

To conclude, from (7) we can see that, for the AWS, *Resp(Z)* can be computed from the entropy *H(Y)* of the automation classification variable, the entropy *H(X)* of the human action selection variable, and their joint entropy *H(X,Y)*.

We need to define the self and mutual distributions of *Y* and *X* to compute their entropies. To do so, we use the simplifying assumptions of a basic equal variance Gaussian SDT model (details are in the appendix): (a) The distributions of the observed values of target and noise are normal with unit variance and means that are *d'* units apart, with the target having the higher mean. (b) The system designers and human operators associate the same cost and benefit values ($V_{FP}$ $V_{TN}$ $V_{FN}$ $V_{TP}$) to the possible outcomes and maximize expected payoff. Hence, we initially assume that system designers and operators share similar incentives and action selection preferences (e.g. they associate a similar high cost, $V_{FP}$, to false engagements of non-targets). (c) The values the human operators associate with different outcomes are independent of the classification results of the automatic module. (d) The information acquisitions by the human and the module, given a certain state of the world, are uncorrelated. This assumption may hold, for example, when the human and the automated module base their information acquisition on different, uncorrelated properties of the state of the world (e.g. optical vs. electromagnetic signatures). (e) The human is rational and has full knowledge of the general characteristics of the automation (its detection sensitivity and cutoff) that determine the automation capabilities. This assumption *does not* mean that the human can supervise the automated module and can determine whether its classifications are correct or wrong.

We first compute the distribution of *Y* (the automated classification variable) and derive its entropy. Substituting $d'_{Automation}$ and $\beta_{Automation}$ into (A7) in the appendix, we can use (A8) to compute the automated module's expected rates of True Positive (*TP*), False Negative (*FN*), False Positive (*FP*) and True Negative (*TN*) which will be denoted by $\tilde{P}_{TP}$, $\tilde{P}_{FN}$, $\tilde{P}_{FP}$, and $\tilde{P}_{TN}$, respectively. Using these rates and the target probability $P_t$, we can compute the distribution of *Y*. For example, the probability that the automated module will classify a random entity as "target" is the probability that the entity is indeed a target and the automation classifies it correctly as such, $P_t\tilde{P}_{TP}$, plus the probability that the entity is actually noise, but the automation falsely classifies it as target, $(1 - P_t)\tilde{P}_{FP}$. In a similar manner, we can compute the probability that the module will classify a random entity as "noise". Table I summarizes the computation results. The computation of the entropy of *Y*, from Table I, is straightforward.

TABLE I
DISTRIBUTION OF *Y*
(CLASSIFICATION RESULTS OF THE AUTOMATED MODULE)

| *Y* (Module classification) | | |
|---|---|---|
| | "Target" | $P_t\tilde{P}_{TP} + (1 - P_t)\tilde{P}_{FP}$ |
| | "Noise" | $P_t\tilde{P}_{FN} + (1 - P_t)\tilde{P}_{TN}$ |

We proceed to compute the joint distribution of *X* (the human action selection variable) and *Y* (the automated classification variable). When aided by the automated module, the human uses two different optimal response criteria, which are derived from the human posterior probability for target or noise, conditioned on the automated module's classification results (see details in the appendix). One criterion is used when the automated module classifies an entity as a target and the other when the automated module classifies an entity as noise. Each response criterion leads to different human values of True Positives, False Negatives, False Positives and True Negatives. We will denote them, respectively, by $P_{TP/"N"}$, $P_{FN/"N"}$, $P_{TN/"N"}$, $P_{FP/"N"}$, when the module indicated that an entity is noise, and by $P_{TP/"T"}$, $P_{FN/"T"}$, $P_{TN/"T"}$, $P_{FP/"T"}$, when the module indicated that an entity is a target. Using these rates, the automation rates and the relative frequency of targets in the environment, we can compute the joint distribution of $X$ and $Y$. For example, the joint probability that the module will classify an entity as "target" and the human will choose to engage it is the probability that the entity is indeed a target, and the conditional probabilities that the automation classifies it correctly as such and the human chooses to engage it, $P_t \tilde{P}_{TP} P_{TP/"T"}$, plus the probability that the entity is actually noise, but the automation falsely classifies it as target, and the human falsely chooses to engage it, $(1 - P_t) \tilde{P}_{FP} P_{FP/"T"}$. In a similar manner, we can compute all other joint probabilities. Table II summarizes the computation of the joint distribution of $X$ and $Y$. The computation of the joint entropy $H(X,Y)$ is straightforward from Table II.

TABLE II
Joint distribution of $X$ (human action selection)
and $Y$ (classification results of the automated module)

|   |   | $X$ (Action selection by the Human) | |
|---|---|---|---|
|   |   | Abort | Engage |
| $Y$ | "Target" | $P_t \tilde{P}_{TP} P_{FN/"T"}$ $+ (1 - P_t) \tilde{P}_{FP} P_{TN/"T"}$ | $P_t \tilde{P}_{TP} P_{TP/"T"}$ $+ (1 - P_t) \tilde{P}_{FP} P_{FP/"T"}$ |
|   | "Noise" | $P_t \tilde{P}_{FN} P_{FN/"N"}$ $+ (1 - P_t) \tilde{P}_{TN} P_{TN/"N"}$ | $P_t \tilde{P}_{FN} P_{TP/"N"}$ $+ (1 - P_t) \tilde{P}_{TN} P_{FP/"N"}$ |

By summing each of the rows in Table II, we can verify that the marginal distribution of $Y$ in Table II is the same as that presented in Table I. For example, we know that $P_{FN/"T"} + P_{TP/"T"} = 1$ and $P_{TN/"T"} + P_{FP/"T"} = 1$, since for each true state of the entity, and a given module's classification of "target", the human either engages or aborts with probability 1. Hence, summing the probabilities in the first row of Table II gives a marginal probability of $P_t \tilde{P}_{TP} + (1 - P_t) \tilde{P}_{FP}$ that the module will classify a random entity as "target", which is equal to the corresponding probability presented in Table I.

In a similar manner, by summing each of the columns in Table II, we can derive the marginal distribution of $X$. Table III presents the results, from which the computation of the entropy of $X$ is straightforward.

TABLE III
Distribution of X (human action selection)

| $X$ | Abort | $P_t(\tilde{P}_{TP} P_{FN/"T"} + \tilde{P}_{FN} P_{FN/"N"}) +$ $(1 - P_t)(\tilde{P}_{FP} P_{TN/"T"} + \tilde{P}_{TN} P_{TN/"N"})$ |
|---|---|---|
|   | Engage | $P_t(\tilde{P}_{TP} P_{TP/"T"} + \tilde{P}_{FN} P_{TP/"N"}) +$ $(1 - P_t)(\tilde{P}_{FP} P_{FP/"T"} + \tilde{P}_{TN} P_{FP/"N"})$ |

D. Quantitative Results

Four variables influence the human action selection process and the resulting human responsibility. These include one environment-related variable (the relative frequency of targets in the environment, $P_t$), the human's and the automation's detection sensitivities ($d'_{Human}$ and $d'_{Automation}$) and the ratio of payoffs the human and the automated module associate with correct and incorrect actions ($V_{ratio} = \frac{V_{TN} - V_{FP}}{V_{TP} - V_{FN}}$), which determines the optimal response criterion (details are in the appendix). Each set of values for these four variables specifies a different combination of environment, automation and human characteristics and relative outcome preferences. This leads to different human and automation rates of True Positives and False Negatives, from which one can compute the distributions on which the human's responsibility calculation in based by using the computation presented in Tables I, II, and III.

*Proposition 1*: The comparative human responsibility, $Resp(Z)$, decreases monotonically in $d'_{Automation}$, and increases monotonically in $d'_{Human}$.

*Proof*: Proof is provided in the Appendix.

Proposition 1 has an intuitive explanation. Under the above assumptions, a human with a given detection sensitivity will rely less on information from less capable automation (in terms of the automation detection sensitivity) than from more capable automation. Thus, the comparative human responsibility increases as the automation capabilities decrease. In addition, for automation with given capabilities, less capable humans will tend to rely more on the automation than would more capable humans. Thus, the comparative human responsibility increases, as the human capabilities increase.

To demonstrate the combined effects of proposition 1, we computed the human responsibility as a function of $d'_{Automation}$ and $d'_{Human}$, each on a scale ranging between .6 (low ability to distinguish between target and noise) and 3 (high ability to distinguish between target and noise). Fig. 4 presents the results. The monotonic properties of $Resp(Z)$ in $d'_{Automation}$ and $d'_{Human}$, are evident.

In the numerical example, presented in Fig.4, we used a target frequency of $P_t = 0.2$, a payoff matrix ratio of $V_{ratio} = 2/3$, and optimal response criteria $\beta_{Automation} = \beta_{Human} = 2.7$. We report below the results of sensitivity analyses of the effects of changes in these values on responsibility outcomes.

*Proposition 2*: Let $R$ denote the detection sensitivities ratio: $R = d'_{Automation}/d'_{Human}$. Suppose that the human and the automation associate the same payoffs with correct and incorrect actions, then $\lim_{R \to \infty} Resp(Z) = 0$ and $\lim_{R \to 0} Resp(Z) = 1$

*Proof:* Proof is provided in the Appendix.

Proposition 2 describes the combined effect of the automation and human detection sensitivities when both have similar preferences and associate the same payoffs with correct and incorrect actions. In this case, when the automation sensitivity is much higher than the human sensitivity (i.e. $R$ is very large) the human responsibility for the output approaches 0, and the human relies mainly on the classifications made by the automated module. In contrast, when the automation sensitivity is much lower than human sensitivity (i.e. $R$ is close to 0), the human responsibility for the output approaches 100%. Here humans rely mainly on their own classification capabilities, ignoring information from the automated module. It is important to note that this proposition implies that even when human sensitivity is not high, the human responsibility may still be high, as long as the automation sensitivity is much lower than that of the human (i.e. as long as $R$ remains low).

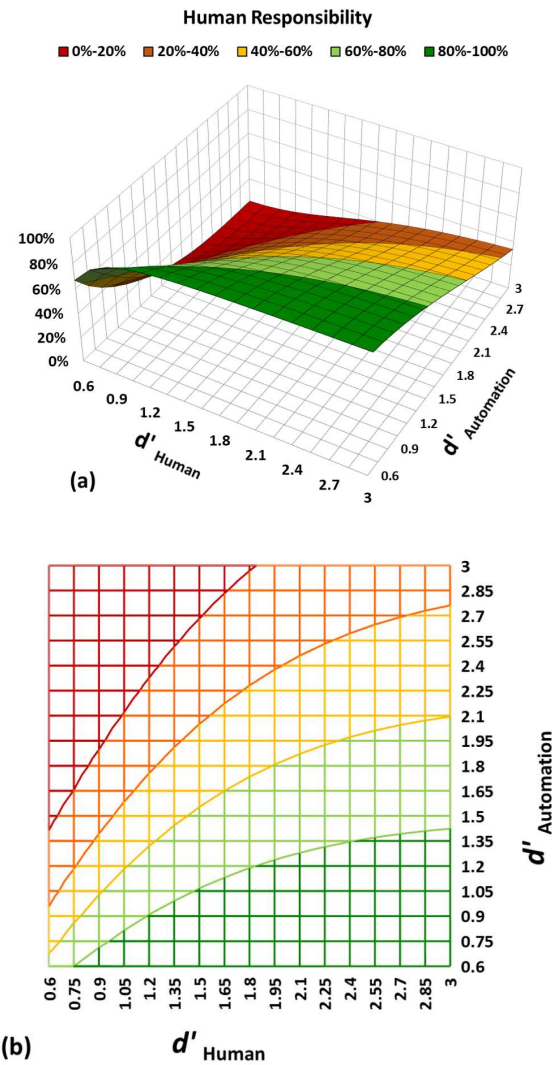

responsibility can have a range of values. This is evident from looking at the main diagonal in Fig. 5(b), which represents a ratio of $R = d'_{Automation}/d'_{Human} = 1$. When $R = 1$, the human responsibility falls into different responsibility regions, depending on the specific values of $d'_{Automation}$ and $d'_{Human}$. In this case, human responsibility is higher when both $d'_{Human}$ and $d'_{Automation}$ are similarly low, compared to when both are similarly high. When $d'_{Human}$ and $d'_{Automation}$ are equally high, the human can still benefit from utilizing the additional information from the automation. The decision will then be based on a similar weighting of the human's own information and the information from the automation, as both are rather accurate, leading to a comparative human responsibility of 40%-60%. However, when both sensitivities are low, the low detection sensitivity of the automation cannot add much to the human decision process. Hence, in this case humans will rely more on their own detection capability, even if it is limited, leading to higher comparative human responsibility for the overall outcomes (60%-80%).

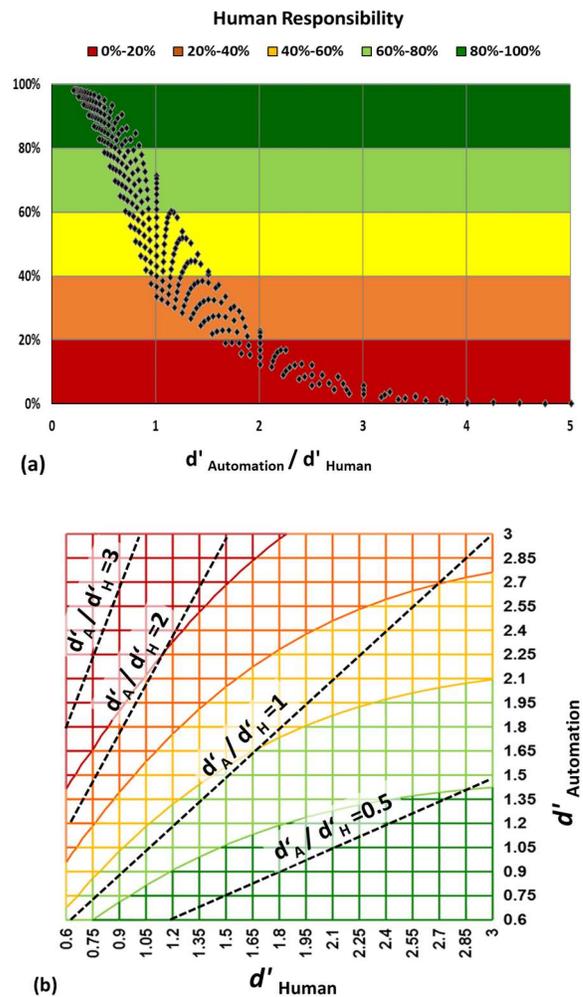

Fig. 4. Three (a) and two (b) dimensional presentation of responsibility values for different combinations of automation and human detection sensitivities ($d'$).

Fig. 5 depicts the human responsibility values as a function of the ratio $R = d'_{Automation}/d'_{Human}$, based on the same assumptions as Fig. 4. Fig. 5(a) shows that the human responsibility may rapidly converge as a function of $R$. When $R$ exceeds 3, the human responsibility is very close to 0, and for $R$ below 1/3, the human is almost fully responsible for the system output. When the automation sensitivity is more than double that of the human (i.e. $R > 2$), the human responsibility drops below 20%.

Fig. 5(a) also demonstrates that when $d'_{Human}$ and $d'_{Automation}$ are similar, so their ratio $R$ is close to 1, the human

Fig.5. Human responsibility as a function of the ratio between the automation and the human detection sensitivities plotted, (a) for different ratios of automation and human detection sensitivities, and (b) on the two-dimensional graph of Fig. 4(a) with dashed lines, representing different examples for fixed sensitivity ratios.

A sensitivity analysis shows that changing the values of the variables that were assumed fixed in Fig. 4 and 5 does not change the above conclusions, as long as both the human and the automation associate the same payoffs with correct and incorrect actions and assume the same relative frequency of targets in the environment.

Matters are different when the human and the automation designers have considerably different preferences, due to different estimates of the costs and benefits associated with different outcomes or of the relative frequency of targets in the environment. In this case, the automation and the human will use different response criteria. Fig. 6 depicts the effects of differences in response criteria on human responsibility for three selected ratios of human and automation sensitivities.

In the first case (Fig. 6a), in which $R=1/3$, the human's detection sensitivity is higher than the low detection sensitivity of the automation. This leads to very high human responsibility, regardless of differences between the human and the automation response criteria.

In the second case (Fig. 6b), in which $R=3$, the automation has a high detection sensitivity, superior to the human's low detection sensitivity. Hence, the human relies mainly on the automation and has low comparative responsibility. However, as is evident from the figure, in this case differences between the human and the automation response criteria have some effect. When the human response criterion differs much from the automation response criterion (is more than 10 times larger or smaller), the human responsibility is considerably higher (increases to 40%-50%) than when the response criteria are similar (human responsibility of less than 10%). Therefore, when there is a large difference between the preferences of the human and the automation, the human will rely less on the automation, even if it has superior detection sensitivity. This has an interesting and non-intuitive interpretation. The response criteria differ when the human and the automation assign different values to possible decision outcomes. In this case, the automation recommendations may disagree with the human incentives, so the human will prefer to rely less on the automation recommendations, even when the automation has better detection capabilities.

In the third case (Fig. 6c) $d'_{Automation}$ is somewhat higher than $d'_{Human}$, and neither value is high. In this case, the effect of differences between the human and the automation response criteria becomes more prominent, because the low detection abilities of the automation cannot compensate for large differences in the preferences.

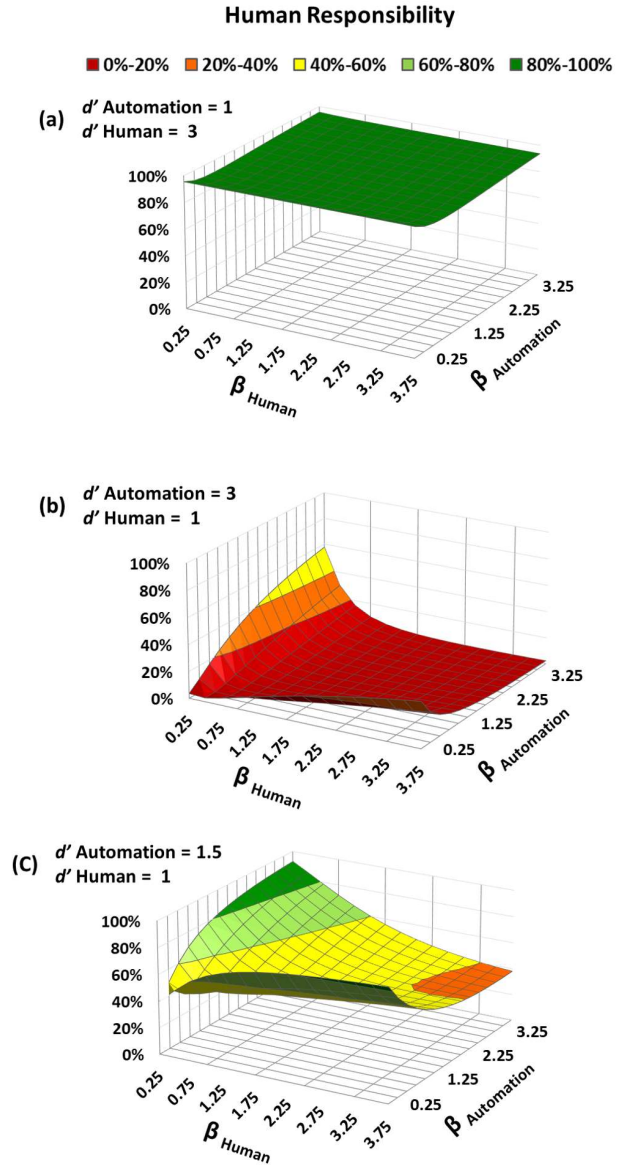

Fig. 6. Human responsibility for different combinations of automation and human response criterion $\beta$. The Figure present the effects of differences in response criteria $\beta$, for three different ratios $R$ (the ratio of automation and human sensitivities): (a) $R=1/3$; (b) $R=3$; (c) $R=1.5$.

IV. DISCUSSION

A. Main Results

By employing information theory measures of entropy and transmission, the ResQu model computes the unique share of human contribution in determining system outcomes. The ResQu responsibility measure enables us to quantify the level of human comparative responsibility in interactions with intelligent systems and advanced automation, and to divide causal responsibility between humans and machines.

Our results demonstrate that the optimal human responsibility depends on human and automation capabilities and on differences between the human and the automation preferences. This optimization is not trivial, as these variables have sometimes contradicting effects. Specifically, the results show that when the human and automation designers have considerably different preferences, a rational human will tend to rely less on the automation, even if the automation has better capabilities, leading to higher comparative human responsibility. When human and automation preferences are similar, a main determinant of human responsibility is the ratio between the automation and human capabilities.

More broadly stated, human causal responsibility in intelligent or automated systems depends on the combined characteristics of the human, the automation, and the operational environment. The combined effects are convoluted. Therefore, human operators may still not be responsible for the system actions and their outcomes, even when important system functions are allocated to the human.

Hence, simplistic demands to keep a human in the loop in order to retain meaningful human control can be misleading and futile. Literally adhering to them may create a mismatch between *role responsibility*, the duties the human operator is accountable for to others, and *causal responsibility*, which is the actual level of human contribution and influence on system outcomes. This may arise from a responsibility-authority double bind, in which the human is assigned a certain role but is not granted the necessary authority to act and control the processes that lead to the outcomes. This can be due to a system design that limits the human's ability to control and take necessary actions to influence the outcomes, or due to probabilistic factors related to the automation and the environment, that limit the effectiveness of the human's action on the combined system outcomes. Thus, simply demanding human involvement does not assure that the human will have a major role in determining the outcomes.

The ResQu model's measure of causal responsibility considers the combined effects of the human-machine system design, the human's role and authority, and probabilistic factors, related to the automation and the environment. Thus, it can be used to measure the level of meaningful human involvement, based on the premise that meaningful human involvement requires the human to have some causal responsibility for the outcomes.

We also show that for some system configurations (when neglecting temporal aspects), both human-on-the-loop and human-in-the-loop levels of control can lead to the same level of comparative human responsibility, because the same information flow model of the combined human-automation system represents both. Thus, the difference between these two systems is not always as substantial as commonly perceived.

### B. Design Implications

When and how should one involve a human in highly intelligent or automated systems? According to our analysis, humans only have significant comparative responsibility when they make unique contributions that supplement or exceed the automated module's capabilities to perform certain functions (e.g., when the human has independent sources of information or is better able to select actions). However, as technologies develop, humans will contribute less to system processes. For instance, future AWS technologies will almost certainly outperform humans in many critical operational tasks, such as the ability to distinguish between combatants and non-combatants, to assess the likelihood of hitting a target or harming civilians, and to decide and act with very short reaction times. When humans will interact with such advanced systems, to which they will contribute very little, they may feel less motivated, or they may attempt to be more involved by interfering more than necessary. Both responses will probably impair the overall system performance. The ResQu model enables system designers to identify such cases in advance and to consider them when evaluating different design alternatives and when planning the human involvement in the system.

With the advent of advanced intelligent systems and automation, with abilities that clearly exceed those of humans in many critical functions, a choice will have to be made. One can progress to fully autonomous systems that keep the human operator out of the loop and abandon the current prevailing demand for a system design with humans in the loop. Alternatively, one can limit the development of autonomous systems and the use of automation. The intermediate option, where systems will be increasingly intelligent, while still keeping the human in the loop, can possibly lead to the inclusion of humans to simply fulfill regulatory requirements, without them having any real impact on system performance.

The current ethical and legal discussion regarding human involvement in intelligent systems and automation should not only focus on the advantages and disadvantages of such systems, but it should also consider the implications of keeping humans in the loop, even when they have little real influence. Falsely claiming that the human is responsible for adverse outcomes, caused by system actions, may expose her or him to unjustified legal liability and to the psychological burden of self-blaming, even when the person actually contributed very little to the outcomes. The ResQu responsibility measure can help in these situations by exposing such anomalies and by providing a new method to quantify the actual human comparative responsibility for the outcomes. This can perhaps lead to a change in the legal treatment of human responsibility in intelligent systems and automation.

### C. Assumptions and Limitations

As a first analytical formulation, the current version of the ResQu model assumes given human and automation capabilities, stationarity, and ignores temporal aspects. Despite these limitations, it can represent interdependencies in complex human-machine systems.

The implementation of the model requires that the modeler builds an information flow model and obtains the values for the underlying distributions of variables, which represent the possible outcomes of human and automation functions and their dependencies. As for other measures of information theory, one needs only the *distributions* of the different variables to calculate the responsibility measure. If the analyzed system is stationary and ergodic, it is possible to infer these distributions from known properties or through empirical observations, taken over time. However, such inference is not possible when systems change and are not stationary (e.g. when there is a learning effect that leads to a change in the level of human involvement over time). In this case one can calculate the responsibility measure repeatedly, at different times, and look if convergence exists (e.g. after performance has stabilized). In addition, the construction of a structured ResQu information flow model, combined with sensitivity analyses of how changes in the input values affect the responsibility measure, will often supply useful insights on the comparative human responsibility. Lastly, when one needs to estimate parameter values, from which the probability distributions are derived (e.g. in SDT models, Cost-Benefit models, etc.), it is often less important to obtain the exact values. It may be enough to estimate the ratios between human and automation values. As we have seen in our application to AWS, the responsibility values can be narrowed down to a limited interval, even if this relative ratio is not exactly known or is hard to assess.

The general ResQu model requires no specific prior assumptions regarding human rationality and behavior. However, when we applied this model to represent human interactions with AWS, using the principles of SDT, we assumed a best-case scenario of perfect rationality on the part of the human, perfect human knowledge of the automation properties and optimal human utilization of information. With these assumptions, the computed human responsibility will be optimal, given the properties of the system. Nevertheless, system designers can use the ResQu model to calculate the sensitivity of the optimal responsibility to those assumptions, for example by analyzing the impact of incomplete human knowledge, such as situations when humans underestimate the automation capabilities or overestimates their own capabilities.

### D. Conclusion and Future Work

The ResQu model is an initial step towards the creation of a comprehensive responsibility model that quantifies human causal responsibility in interactions with intelligent systems and advanced automation. The model can serve as an additional tool in the analysis of system design alternatives and policy decisions regarding human responsibility.

Future work should expand the model, enabling it to deal with temporal effects, such as the time required to make a decision and its effects on the human's tendency to rely on the automation. To do so, the information theoretical framework we present here should be expanded to address temporal aspects by evaluating not just transmitted information, but also information transmission rates and by defining a responsibility measure that also considers human channel capacity constraints.

Future work should also test the predictive ability of the ResQu model by comparing the computed theoretical values to actual human performance, and by tying it to existing empirical research on human-automation interaction. A first empirical analysis of the ResQu model demonstrated that the model is not merely an abstract theoretical model, but it can also serve as a descriptive model, that allows us to predict the actual responsibility users take on when using a system [90].

Lastly, future work should analyze the sensitivity of the ResQu model's responsibility estimates to different measurement errors of the input variables and their dependencies. Such analyses can help to identify the important variables that practitioners should focus on to obtain accurate estimates when applying the model.

## V. APPENDIX

This appendix presents the basic Signal Detection Theory (SDT) concepts and formulas we used to model the probabilistic nature of the AWS and to perform the numerical calculations, leading to the results presented in the manuscript.

The basic SDT model describes a system with a single sensor, observing an environment with only two possible entities: Target+Noise (referred to as Target) and Noise alone (referred to as Noise) that occur with probability $P_t$ and $1- P_t$, respectively. Both entities can be measured by a single observable parameter, which transforms the data into a scale value. The distributions of values of the observed characteristics for target and noise entities differ (with targets usually assumed to have a larger mean value than noise), which allows some discrimination between the two types of entities. However, the distributions overlap, so when a certain value is observed, there is uncertainty whether the entity is indeed a target, or whether it is actually noise. We use Gaussian distributions for the example, but the model does not depend on the assumptions regarding the distributions.

The sensor is required to identify and engage targets and to prevent engagement of noise. This binary decision is categorized as Engage or Abort. The responses are the outcomes of the decision process and can be categorized as True Positive (*TP*) when a target is present and the response is to engage, False Negative (*FN*) when a target is present and the response is not to engage, True Negative (*TN*) when no target is present and the response is not to engage, and False Positive (*FP*) when no target is present and the response is to engage. Table A.1 summarizes the classification of human responses.

TABLE A.1
CLASSIFICATION OF HUMAN RESPONSES USING SDT

|  |  | Human Response | |
|---|---|---|---|
|  |  | Engage | Abort |
| Actual Environment state | Target | True Positive (TP) | False Negative (FN) |
|  | Noise | False Positive (FP) | True Negative (TN) |

Signal detection theory differentiates between the detection sensitivity of a sensor and its response bias. The *detection sensitivity* ($d'$) is the sensor's ability to distinguish between target and noise. This is represented by the shift of the signal probability density function, compared to the noise probability density function. When $d'=0$, the sensor is unable to distinguish between target and noise. As $d'$ increases, the ability to distinguish between the two entities increases.

For every value of the observed parameter, one can compute the likelihoods of observing the value under the target distribution or the noise distributions. We assume a threshold likelihood ratio. This threshold is called the *response criterion* ($\beta$). The response criterion represents the sensor's tendency to favor one response over the other. The value of the observed parameter at the threshold is the *cutoff point* (*C*). When the observed value is below the cutoff point, the observation is classified as noise, and it is above the cutoff point, it is classified as a target. The values of $d'$ and $\beta$ determine the probabilities of the four possible outcomes (*TP, FN, FP,* and *TN*) as presented in Fig. A.1.

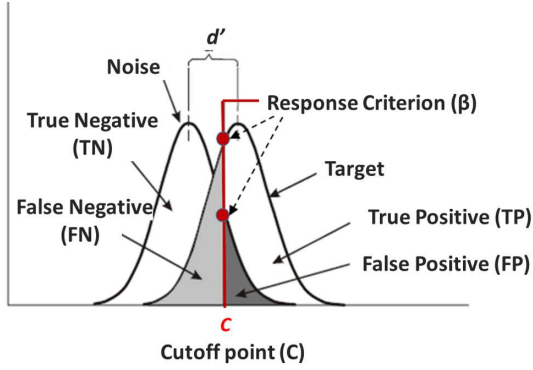

Fig. A.1. The basic SDT model for Gaussian distributions, with probability density functions for target and noise, detection sensitivity ($d'$), response criterion ($\beta$), cutoff point (*C*) and probabilities of possible outcomes.

The distributions of noise and target over the values of the observable variable are denoted by $E_n, E_t$. In the basic normal, equal variance SDT model we have:

$$E_n \sim N(\mu_n, \sigma_n^2) \quad E_t \sim N(\mu_t, \sigma_t^2) \quad \text{(A1)}$$

$$\mu_n = -.5d' \quad \mu_t = .5d' \quad \sigma_n^2 = \sigma_t^2 = 1 \quad \text{(A2)}$$

In this case:

$$E_n \sim N(-.5d', 1) \quad E_t \sim N(.5d', 1) \quad \text{(A3)}$$

$$d' = \mu_t - \mu_n \quad \text{(A4)}$$

$$\beta = \frac{f_t(c)}{f_n(c)} \quad \text{(A5)}$$

$$\ln\beta = \ln\frac{f_t(c)}{f_n(c)} = \ln f_t(c) - \ln f_n(c) = d'c \quad \text{(A6)}$$

$$c = \frac{\ln\beta}{d'} \quad \text{(A7)}$$

The probability for different outcomes can be calculated as (see Fig A.1):

$$P(TP) = P(E_t > c) \quad P(FN) = P(E_t \leq c)$$
$$P(FP) = P(E_n > c) \quad P(TN) = P(E_n \leq c) \quad \text{(A8)}$$

Assume that there are cost-benefit values associated with each outcome: $V_{FP} V_{TN} V_{FN} V_{TP}$ (where: $V_{FP}$ and $V_{FN}$ are negative costs, $V_{TP}$ and $V_{TN}$ are positive benefits). It can be shown that the optimal response criterion, $\beta^*$, which maximizes the expected value is:

$$\beta^* = \frac{1-P_t}{P_t}\frac{V_{TN}-V_{FP}}{V_{TP}-V_{FN}} \quad \text{(A9)}$$

Where $P_t$ is the target probability and $1-P_t$ is the noise probability. We denote the ratio of the cost-benefit values as:

$$V_{ratio} = \frac{V_{TN}-V_{FP}}{V_{TP}-V_{FN}} \quad \text{(A10)}$$

To conclude, under the above assumptions, if we know the probability of a signal in the environment ($P_s$), the sensor's sensitivity ($d'$), and the ratio of the cost-benefit values ($V_{ratio}$), we can use the above formulas to calculate an optimal response criterion ($\beta^*$) that maximizes the expected value. We can also compute the True Positive and False Positive rates.

The human and the automated module may have different *detection sensitivities* ($d'_{Human}$ and $d'_{Automation}$), leading to different capabilities to classify whether a given entity is a legitimate target or noise. In addition, the human and the automation may also differ in the threshold value above which they classify an entity as a target (with response criteria $\beta_{Human}$ and $\beta_{Automation}$, respectively).

We next examine the case where the human detection is aided by the automated module, which produces an alert when it identifies a suspected target. Assume that the automated module has a detection sensitivity $d'_{Automation}$ and a response criterion $\beta_{Automation}$, which are known to the human. Denote the module's rates of True Positives by $\tilde{P}_{TP}$ and False Positives by $\tilde{P}_{FP}$. Using Bayes' law, the human can use these probabilities to update the prior probability of signal in the environment, according the automation classification results.

Denote by $\hat{P}_{t/"T"}$, the human posterior probability for target, when the automated module classifies an entity as target.

$$\hat{P}_{t/"T"} = \frac{P_t\tilde{P}_{TP}}{P_t\tilde{P}_{TP}+(1-P_t)\tilde{P}_{FP}} \quad \text{(A11)}$$

The human uses $\hat{P}_{t/"T"}$, instead of $P_t$ in (A9), to compute the optimal response criterion that maximizes the expected value, given that the module has classified an entity as a target.

Denote by $\hat{P}_{t/"N"}$, the human posterior probability for target, when the automated module classifies an entity as noise.

$$\hat{P}_{t/"N"} = \frac{P_t\tilde{P}_{FN}}{P_t\tilde{P}_{FN}+(1-P_t)\tilde{P}_{TN}} = \frac{P_t(1-\tilde{P}_{TP})}{P_t(1-\tilde{P}_{TP})+(1-P_t)(1-\tilde{P}_{FP})} \quad \text{(A12)}$$

In the same manner, the human uses $\hat{P}_{t/"N"}$ (A9) to compute the optimal response criterion that maximizes the expected value, given that the module has classified an entity as a noise.

Thus, when aided by an automated module, the human uses two different response criteria, one when the automated module classifies an entity as a target and the other, when the automated module classifies an entity as noise. The human cutoff point when the entity was classified as target by the automation is smaller than when it was classified as noise. Fig. A.2 presents the two cutoff points, when human detection is aided by an automated module.

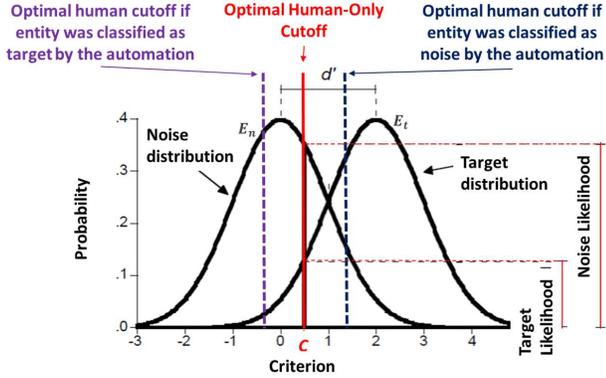

Fig. A.2. SDT model when human detection is aided by an automated module. There are two cutoff points according to the module's classification results.

By adjusting the threshold according to the classification of the automated module, the human increases the probability of distinguishing between target and noise.

Denote by $d'_{effective}$ the combined sensitivity of such a system. This is essentially the sensitivity equivalent to a single Gaussian SDT detector that has the same level of performance as the combined tandem human-automation system. By definition, $d'_{effective}$ is greater or equal than $d'_{Human}$ and $d'_{Automation}$.

Assume that (a) the distributions of the observed values of target and noise are normal with unit variance; (b) the cost and benefit values $V_{FP}\ V_{TN}\ V_{FN}\ V_{TP}$ are the same for the human and the automated module; (c) the cost and benefit values for the human are independent of the classification results of the automatic module; (d) the initial information the human and the module have about a the state of the world are uncorrelated.

Pollack and Madans [91] have shown that under the above simplifying assumptions the maximum value of $d'_{effective}$, when the detectors preserve continuous information and an optimal decision rule is employed, will be equal to

$$d'_{max} = \sqrt{{d'_{Human}}^2 + {d'_{Automation}}^2} \quad (A13)$$

In our system, there is some loss of information, since the automation provides only binary information rather than continuous, so in most cases $d'_{effective}$ will be lower than $d'_{max}$.

$$d'_{effective} \leq \sqrt{{d'_{Human}}^2 + {d'_{Automation}}^2} \quad (A14)$$

$$d'_{Human} \leq d'_{effective} \qquad d'_{Automation} \leq d'_{effective}$$

*Lemma 1:* $Resp(Z)$ is monotonically increasing in $d'_{Human}$.
*Proof:* Assume that $d'_{Human}$ increases, and all other variables remain fixed. In particular, since $d'_{Automation}$ remains fixed, so do the human posterior probabilities for target $\hat{P}_{t/"T"}$ and $\hat{P}_{t/"N"}$, presented in equations (A11) and (A12). Denote by $\beta^*_{"T"}$ and $\beta^*_{"N"}$, respectively, the optimal human response criteria that maximizes the expected value, given that the module has classified an entity as a target or noise.

$$\beta^*_{"T"} = \frac{1-\hat{P}_{t/"T"}}{\hat{P}_{t/"T"}} \frac{V_{TN}-V_{FP}}{V_{TP}-V_{FN}} \qquad \beta^*_{"N"} = \frac{1-\hat{P}_{t/"N"}}{\hat{P}_{t/"N"}} \frac{V_{TN}-V_{FP}}{V_{TP}-V_{FN}} \quad (A15)$$

These two criteria remain fixed as $d'_{Human}$ increases. For each of them, there is a corresponding cutoff point that can be derived using equation (A7). As $d'_{Human}$ increases, it follows from (A7) that the weight the human gives to the automation decreases monotonically to zero, so the two corresponding cutoff points are moving towards each other, approaching 0. This means that as $d'_{Human}$ increases, when selecting an action, the human assigns more weight to $d'_{Human}$ and less weight to the automation classification results.

In terms of information theory, this means that as $d'_{Human}$ increases, the human action selection variable $X$ depends less on the automation classification variable, $Y$. Hence their mutual information $I(X;Y)$ decreases monotonically, $H(X/Y)$ increases monotonically, and so does $H(X/Y)/H(X)$. From equations (6) we can conclude that $Resp(Z)$ is monotonically increasing in $d'_{Human}$ □

*Lemma 2:* $Resp(Z)$ is monotonically decreasing in $d'_{Automation}$
*Proof:* Assume that $d'_{Automation}$ increases, and all other variables remain fixed. As $d'_{Automation}$ increases, both $\tilde{P}_{TP} \rightarrow 1$ and $\tilde{P}_{TN} \rightarrow 1$ monotonically.

We first examine the case when the automated module classifies an entity as a target. Here, the human uses $\hat{P}_{t/"T"}$ to compute the optimal response criterion, instead of $P_t$ in (A9). From (A11) we get $\hat{P}_{t/"T"} \rightarrow 1$ as $\tilde{P}_{TP} \rightarrow 1$. The increase in $\hat{P}_{t/"T"}$ lowers the human response criterion, and from (A7) this lowers the corresponding human cutoff point, so $P_{TP/"T"} \rightarrow 1$. Denote by $P_{Engage/"T"}$ the probability that the human will choose to engage an entity, given that it was classified as a target by the automation.

$$P_{Engage/"T"} = \hat{P}_{t/"T"} P_{TP/"T"} + (1-\hat{P}_{t/"T"}) P_{FP/"T"} \quad (A16)$$

From the above, when $d'_{Automation}$ increases, $P_{Engage/"T"}$ monotonically increases to 1. In a similar manner, it can be shown that when $d'_{Automation}$ increases and an entity is classified as noise by the automation, $P_{Engage/"N"}$ decreases monotonically to 0.

Therefore, as $d'_{Automation}$ increases, there is a higher probability that the human will act according to the automation classification, engaging an entity the automated module classified as target and not engaging an entity it classified as noise.

In terms of information theory, this means that as $d'_{Automation}$ increases, the automation classification variable, $Y$, provides more information to the human action selection variable $X$, monotonically reducing $H(X/Y)$, the remaining uncertainty about $X$ when $Y$ is known, towards zero. Denote by $T$ a Bernoulli variable that corresponds to the prevalence of targets in the environment. As $d'_{Automation}$ increases, the distribution of $Y$ approaches the distribution of $T$ so $H(Y)$ approaches a fixed known value $H(T) \in (0,1)$. In addition, as $d'_{Automation}$ increases, the distribution of $X$ approaches the distribution $Y$, so $H(X)$ also approaches $H(T)$. From equations (6) we get

$$Resp(Z) = \frac{H(X/Y)}{H(X)} \to \frac{0}{H(T)} = 0 \quad (A17)$$

Thus, we can conclude that *Resp(Z)* monotonically decreases in $d'_{Automation}$ □

*Proposition 1*: The comparative human responsibility *Resp(Z)* is monotonically decreasing in $d'_{Automatio}$, and monotonically increasing in $d'_{Human}$.

*Proof:* Proof is immediate from Lemma 1 and Lemma 2. □

*Proposition 2*: Let $R$ denote the detection sensitivities ratio: $R = d'_{Automation} / d'_{Human}$. Suppose that the human and the automation associate the same payoffs with correct and incorrect actions, then $\lim_{R \to \infty} Resp(Z) = 0$ and $\lim_{R \to 0} Resp(Z) = 1$

*Proof:* Under condition of the proposition and our model assumptions, equation (A14) holds. Using $R = d'_{Automation} / d'_{Human}$:

$$d'_{Automation} \leq d'_{effective} \leq \sqrt{\frac{d'_{Automation}{}^2}{R^2} + d'_{Automation}{}^2} \quad (A18)$$

Consequently $\lim_{R \to \infty} d'_{effective} = d'_{Automation}$. This means that when both human and automation associate the same payoffs with correct and incorrect actions, and $d'_{Automation}$ is much larger than $d'_{Human}$, a rational human will base the action selection decision primarily upon the results of the automation classification. Therefore, when $R \to \infty$ the human action selection variable, $X$, will be fully determined by the automation classification variable, $Y$, and thus will have the same distribution as $Y$. In terms of entropy this means

$$\lim_{R \to \infty} H(X/Y) = 0 \qquad \lim_{R \to \infty} H(X) = H(Y) \quad (A19)$$

From equations (6) we have

$$\lim_{R \to \infty} Resp(Z) = \lim_{R \to \infty} \frac{H(X/Y)}{H(X)} = \frac{0}{H(Y)} = 0 \quad (A20)$$

The proof for $R \to 0$ is analogical. In this case, we have:

$$d'_{Human} \leq d'_{effective} \leq \sqrt{d'_{Human}{}^2 + R^2 d'_{Human}{}^2} \quad (A21)$$

Thus, $\lim_{R \to 0} d'_{effective} = d'_{Human}$. This means that when $d'_{Human}$ is much larger than $d'_{Automation}$, and both human and automation associate the same payoffs with correct and incorrect actions, rational humans will base the action selection decision primarily on their own detection capabilities. Therefore, when $R \to 0$ the human action selection variable, $X$, will be independent from the automation classification variable, $Y$. In terms of entropy this means:

$$\lim_{R \to 0} H(X/Y) = H(X) \quad (A22)$$

$$\lim_{R \to 0} Resp(Z) = \lim_{R \to 0} \frac{H(X/Y)}{H(X)} = \frac{H(X)}{H(X)} = 1 \quad (A23)$$

This completes the proof. □

REFERENCES

[1] M. Bergsten and J. Sandahl, "Algorithmic trading in the foreign exchange market," *Sveriges Riksbank Economic Review, (1),* pp. 31, 2013.
[2] T. Hendershott, C. M. Jones and A. J. Menkveld, "Does Algorithmic Trading Improve Liquidity?" *J. Finance,* vol. 66, (1), pp. 1-33, 2011. doi: 10.1111/j.1540-6261.2010.01624.x.
[3] L. Da-Yin, "Automation and integration in semiconductor manufacturing, semiconductor technologies," Jan Grym (Ed.), *InTech,* pp.39-56. 2010. Available: https://www.intechopen.com /books/ semiconductor-technologies/automation-and-integration-in-semiconductor-manufacturing.
[4] K. Doi, "Computer-aided diagnosis in medical imaging: historical review, current status and future potential," *Comput. Med. Imaging Graphics,* vol. 31, *(4),* pp. 198-211, 2007.
[5] R. M. Rangayyan, F. J. Ayres and J. L. Desautels, "A review of computer-aided diagnosis of breast cancer: Toward the detection of subtle signs," *Journal of the Franklin Institute,* vol. 344, *(3),* pp. 312-348, 2007.
[6] C. E. Billings, *Aviation Automation*. Mahwah, NJ: Erlbaum, USA, 1997.
[7] A. R. Pritchett, "Aviation Automation: General Perspectives and Specific Guidance for the Design of Modes and Alerts," *Reviews of Human Factors and Ergonomics,* vol. 5, *(1),* pp. 82-113, 2009. doi: 10.1518/155723409X448026.
[8] T. Litman, "Autonomous vehicle implementation predictions" *Victoria Transport Policy Institute*, pp. 1-24, 2017 Available: http://www.vtpi.org/avip.pdf.
[9] T. Luettel, M. Himmelsbach and H. Wuensche, "Autonomous Ground Vehicles Concepts and a Path to the Future," *Proc IEEE*, vol. 100, pp. 1831-1839, 2012. doi: 10.1109/JPROC.2012.2189803.
[10] H. L. A. Hart and T. Honor, *Causation in the Law*. Oxford University Press, Oxford, 1985.
[11] H. L. A. Hart, P*unishment and Responsibility: Essays in the Philosophy of Law*. Oxford University Press, Oxford, 2008.
[12] N. A. Vincent, "A structured taxonomy of responsibility concepts," in *Moral Responsibility*, N.A. Vincent, I. van de Poel, and J. Hoven, eds., Springer, Netherlands, pp. 15-35, 2011.
[13] M. D. Alicke, D. R. Mandel, D. J. Hilton, T. Gerstenberg and D. A. Lagnado, "Causal conceptions in social explanation and moral evaluation: A historical tour," *Perspectives on Psychological Science,* vol. 10, *(6),* pp. 790-812, 2015.
[14] F. Cushman, "Crime and punishment: Distinguishing the roles of causal and intentional analyses in moral judgment," *Cognition,* vol. 108, *(2),* pp. 353-380, 2008.
[15] B. F. Malle, S. Guglielmo and A. E. Monroe, "A theory of blame," *Psychological Inquiry,* vol. 25, *(2),* pp. 147-186, 2014.
[16] R. Rogers, M. D. Alicke, S. G. Taylor, D. Rose, T. L. Davis and D. Bloom, "Causal deviance and the ascription of intent and blame," P*hilosophical Psychology,* vol. 32, *(3),* pp.404-427, 2019.

[17] K. G. Shaver, *The Attribution of Blame: Causality, Responsibility, and Blameworthiness.* Springer Science & Business Media, 2012.
[18] M. S. Moore, *Causation and Responsibility: An Essay in Law, Morals, and Metaphysics.* Oxford University Press on Demand, 2009.
[19] S. Steel, *Proof of Causation in Tort Law.* Cambridge University Press, 2015.
[20] R. W. Wright, "Causation in tort law," *California Law Review,* vol. 73, pp. 1735, 1985.
[21] R. Parasuraman, T. B. Sheridan and C. D. Wickens, "A model for types and levels of human interaction with automation," *IEEE Transactions on Systems Man and Cybernetics. A. Systems and Humans*, vol. 30, (3), pp. 286-297, 2000. doi: 10.1109/3468.844354.
[22] D. A. Abbink, T. Carlson, M. Mulder, De Winter, J. C. F., F. Aminravan, T. L. Gibo and E. R. Boer, "A topology of shared control systems-finding common ground in diversity," *IEEE Transactions on Human- Machine Systems*, vol. 48, (5), pp. 509-525, 2018. doi: 0.1109/THMS.2018.2791570.
[23] M. Johnson, J. M. Bradshaw, P. J. Feltovich, C. M. Jonker, B. Van Riemsdijk and M. Sierhuis, "The fundamental principle of coactive design: Interdependence must shape autonomy," in *International Workshop on Coordination, Organizations, Institutions, and Norms in Agent Systems*, pp. 172-191, 2010.
[24] M. Johnson, J. M. Bradshaw, P. J. Feltovich, C. M. Jonker, M. B. Van Riemsdijk and M. Sierhuis, "Coactive design: Designing support for interdependence in joint activity," *Journal of Human-Robot Interaction*, vol. 3, (1), pp. 43-69, 2014.
[25] D. Castelvecchi, "Can we open the black box of AI?" *Nature News,* vol. 538, (7623), pp. 20, 2016.
[26] P. Scharre, *Autonomous Weapons and Operational Risk.* Center for a New American Security Washington, DC, 2016.
[27] D. G. Johnson and T. M. Powers, "Computer systems and responsibility: A normative look at technological complexity," *Ethics and Information Technology,* vol. 7, (2), pp. 99-107, 2005.
[28] M. Coeckelbergh, "Moral responsibility, technology, and experiences of the tragic: From Kierkegaard to offshore engineering," *Science and engineering ethics,* vol. 18, (1), pp. 35-48, 2012.
[29] R. D. Cooter and T. S. Ulen, "An economic case for comparative negligence," *New York University Law Review*, vol. 61, pp. 1067, 1986.
[30] J. V. Pinto, "Comparative Responsibility-An Idea Whose Time Has Come," *Insurance Counsel Journal,* vol. 45, pp. 115, 1978.
[31] D. C. Sobelsohn, "Comparing Fault," *Indiana Law Journal,* vol. 60, pp. 413-462, 1984.
[32] D. D. Woods, "Cognitive technologies: The design of joint human-machine cognitive systems," *AI Magazine,* vol. 6, (4), pp. 86, 1985.
[33] D. D. Woods, "Conflicts between learning and accountability in patient safety," *DePaul Law Review,* vol. 54, pp. 485-502, 2004.
[34] A. Matthias, "The responsibility gap: Ascribing responsibility for the actions of learning automata," *Ethics and information technology,* vol. 6, (3), pp. 175-183, 2004. doi: 10.1007/s10676-004-3422-1.
[35] R. Sparrow, "Killer robots," *Journal of Applied Philosophy,* vol. 24, (1), pp. 62-77, 2007.
[36] D. G. Johnson, "Technology with No Human Responsibility?" *Journal of Business Ethics,* vol. 127, (4), pp. 707, 2014. doi: 10.1007/s10551-014-2180-1.
[37] H. Chockler and J. Y. Halpern, "Responsibility and blame: A structural model approach," *Journal of Artificial Intelligence Research,* vol. 22, pp. 93-115, 2004.
[38] T. Gerstenberg and D. A. Lagnado, "Spreading the blame: The allocation of responsibility amongst multiple agents," *Cognition,* vol. 115, (1), pp. 166-171, 2010.
[39] D. A. Lagnado, T. Gerstenberg and R. Zultan, "Causal responsibility and counterfactuals," *Cognitive Science,* vol. 37, (6), pp. 1036-1073, 2013.
[40] A. F. Langenhoff, A. Wiegmann, J. Y. Halpern, J. B. Tenenbaum and T. Gerstenberg, "Predicting responsibility judgments from dispositional inferences and causal attributions, Working Paper, 2019.
[41] R. Zultan, T. Gerstenberg and D. A. Lagnado, "Finding fault: causality and counterfactuals in group attributions." *Cognition,* vol. 125, (3), pp. 429-440, 2012.
[42] M. L. Cummings, "Lethal Autonomous Weapons: Meaningful human control or meaningful human certification?" *IEEE Technology and Society Magazine,* vol. 38, (4), pp. 20-26, 2019.
[43] L. Righetti, Q. Pham, R. Madhavan and R. Chatila, "Lethal Autonomous Weapon Systems [Ethical, Legal, and Societal Issues]," *IEEE Robotics & Automation Magazine,* vol. 25, (1), pp. 123-126, 2018.
[44] ICRC, "Autonomous weapon systems: Implications of increasing autonomy in the critical functions of weapons," in *Expert Meeting of International Committee of the Red Cross,* pp. 1-94, 2016. Available: http://icrcndresourcecentre.org/wp-content/uploads/2017/11/4283_002_Autonomus-Weapon-Systems_WEB.pdf
[45] A. Wyatt, "Charting great power progress toward a lethal autonomous weapon system demonstration point," *Defence Studies,* pp. 1-20, 2019.
[46] R. Crootof, "The Killer Robots Are Here: Legal and Policy Implications," *Cardozo Law Review*, vol. 36, (5), 2015.
[47] B. Docherty, R. A. Althaus, A. Brinkman, C. Jones and R. B. Skipper, "Losing Humanity: The Case Against Killer Robots," *Science and Engineering Ethics,* vol. 20, (1), 2012.
[48] R. Sparrow, "Predators or plowshares? Arms control of robotic weapons," *IEEE Technology and Society Magazine*, vol. 28, (1), pp. 25-29, 2009.
[49] M. L. Cummings, "Automation and Accountability in Decision Support System Interface Design." *Journal of Technology Studies,* vol. 32, (1), pp. 23-31, 2006.
[50] A. Gerdes, "Lethal autonomous weapon systems and responsibility gaps," *Philosophy Study,* vol. 8, (5), pp. 231-239, 2018.
[51] P. Asaro, "On banning autonomous weapon systems: human rights, automation, and the dehumanization of lethal decision-making," *International Review of the Red Cross,* vol. 94, (886), pp. 687-709, 2012.
[52] B. L. Docherty, *Mind the Gap: The Lack of Accountability for Killer Robots.* Human Rights Watch, 2015.
[53] S S. Goose, "The case for banning killer robots," *Communications of the ACM,* vol. 58, (12), pp. 43-45, 2015. doi: 10.1145/2835963
[54] A. Guersenzvaig, "Autonomous Weapon Systems: Failing the Principle of Discrimination," *IEEE Technology and Society Magazine,* vol. 37, (1), pp. 55-61, 2018.
[55] A. Hauptman, "Autonomous Weapons and the Law of Armed Conflict," *Military Law Review,* vol. 218, pp. 170-195, 2013.
[56] T. Hellström, "On the moral responsibility of military robots," *Ethics and Information Technology*, vol. 15, (2), pp. 99-107, 2013.
[57] M. Noorman and D. G. Johnson, "Negotiating autonomy and responsibility in military robots," *Ethics and Information Technology,* vol. 16, (1), pp. 51-62, 2014.
[58] M. Noorman, "Responsibility practices and unmanned military technologies," *Sci. Eng. Ethics,* vol. 20, (3), pp. 809-826, 2014.
[59] N. Sharkey, "Saying 'no!'to lethal autonomous targeting," *Journal of Military Ethics,* vol. 9, (4), pp. 369-383, 2010.
[60] J. I. Walsh, "Political accountability and autonomous weapons," *Research & Politics*, vol. 2, (4), pp. 1-6, 2015.
[61] USDD, "Directive 3000.09: Autonomy in Weapon Systems," U*nited States of America: Department of Defense*, pp. 1-15, 2012. Available: http://www.esd.whs.mil/Portals/54/Documents/DD/issuances/ dodd/ 300009p.pdf.
[62] I ICRC, "Autonomous weapon systems: Technical, military, legal and humanitarian aspects," *International Committee of the Red Cross*. pp.1-202, 2014. Available: https://www.icrc.org/en/download/ file/ 1707/4221-002-autonomous-weapons-systems-full-report.pdf.
[63] UNIDIR, "The weaponization of increasingly autonomous technologies: Considering how meaningful human control might move the discussion forward," *United Nations Institute for Disarmament Research*, pp.1-9. 2014. Available:http://www.unidir.ch/files/publications/pdfs/considering-how-meaningful-human-control-might-move-the-discussion-forward-en-615.pdf.
[64] M. Horowitz and P. Scharre, *Meaningful Human Control in Weapon Systems: A Primer.* Center for a New American Security, 2015.
[65] F. S. de Sio and d. H. van, "Meaningful human control over autonomous systems: A philosophical account," *Frontiers in Robotics and AI*, vol. 5, pp. 1-14 , 2018. doi: 10.3389/frobt.2018.00015.
[66] F. Ficuciello, G. Tamburrini, A. Arezzo, L. Villani and B. Siciliano, "Autonomy in surgical robots and its meaningful human control," *Paladyn, Journal of Behavioral Robotics,* vol. 10, (1), pp. 30-43, 2019.
[67] G. Mecacci and F. S. de Sio, "Meaningful human control as reason-responsiveness: the case of dual-mode vehicles," *Ethics and Information Technology,* pp. 1-13, 2019.
[68] F. Santoni de Sio and J. Van den Hoven, "Meaningful human control over autonomous systems: a philosophical account," *Frontiers in Robotics and AI,* vol. 5, pp. 15, 2018.

[69] A. R. Pritchett, S. Y. Kim and K. M. Feigh, "Measuring Human-Automation Function Allocation," *Journal of Cognitive Engineering and Decision Making*, vol. 8, *(1)*, pp. 52-77, 2014. doi: 10.1177/1555343413490166.

[70] M. C. Elish, "Moral Crumple Zones: Cautionary Tales in Human-Robot Interaction," *We Robot 2016*, 2016.

[71] M. C. Elish and T. Hwang, "Praise the Machine! Punish the Human! The Contradictory History of Accountability in Automated Aviation," *Intelligence & Autonomy Working Paper*, Data & Society Research Institute 2015.

[72] M. C. Canellas and R. A. Haga, "Toward meaningful human control of autonomous weapons systems through function allocation,", *IEEE International Symposium on Technology and Society (ISTAS 2015)*, pp. 1-7, 2015.

[73] R. C. Conant, "Laws of information which govern systems," *IEEE transactions on systems, man, and cybernetics*,*(4)*, pp. 240-255, 1976.

[74] C. E. Shannon, "A mathematical theory of communication," *Bell system technical journal* vol. 27, *(3)*, pp. 379-423, 1948.

[75] T. M. Cover and J. A. Thomas, *Elements of Information Theory.* John Wiley & Sons, New York, USA, 2012.

[76] H. Theil, "On the Estimation of Relationships Involving Qualitative Variables," *American Journal of Sociology,* vol. 76, *(1)*, pp. 103-154, 1970. doi: 10.1086/224909.

[77] H. Theil, *Statistical Decomposition Analysis : With Applications in the Social and Administrative Sciences.* Amsterdam, North-Holland Pub. Co, 1972.

[78] J. Meyer, "Effects of warning validity and proximity on responses to warnings," *Human Factors,* vol. 43, *(4),* pp. 563-572, 2001.

[79] J. Meyer, "Conceptual issues in the study of dynamic hazard warnings," *Human Factors,* vol. 46, *(2),* pp. 196-204, 2004.

[80] G. Vashitz, J. Meyer, Y. Parmet, R. Peleg, D. Goldfarb, A. Porath and H. Gilutz, "Defining and measuring physicians' responses to clinical reminders," J*ournal of Biomedical Informatics,* vol. 42, *(2),* pp. 317-326, 2009.

[81] K. M. Feigh and A. R. Pritchett, "Requirements for effective function allocation: A critical review," *Journal of Cognitive Engineering and Decision Making,* vol. 8, *(1),* pp. 23-32, 2014.

[82] M. Canellas and R. Haga, "Lost in Translation: Building a Common Language for Regulating Autonomous Weapons," *IEEE Technology and Society Magazine,* vol. 35, *(3),* pp. 50-58, 2016.

[83] D. Eskins and W. H. Sanders, "The multiple-asymmetric-utility system model: A framework for modeling cyber-human systems," in *2011 Eighth International Conference on Quantitative Evaluation of SysTems,* 2011, pp. 233-242.

[84] J. Cámara, G. A. Moreno and D. Garlan, "Reasoning about human participation in self-adaptive systems," in *Proceedings of the 10th International Symposium on Software Engineering for Adaptive and Self-Managing Systems,* 2015, pp. 146-156.

[85] J. Cámara, D. Garlan, G. A. Moreno and B. Schmerl, "Evaluating trade-offs of human involvement in self-adaptive systems," in *Managing Trade-Offs in Adaptable Software Architectures,* Morgan Kaufmann, pp. 155-180, 2017.

[86] M. Gil, M. Albert, J. Fons and V. Pelechano, "Designing human-in-the-loop autonomous Cyber-Physical Systems," *International Journal of Human-Computer Studies,* vol. 130, pp. 21-39, 2019.

[87] D. M. Green and J. A. Swets, *Signal Detection Theory and Psychophysics.* New York, USA, Wiley, 1966.

[88] N. A. Macmillan and C. D. Creelman, *Detection Theory: A User's Guide.* New York, NY, USA, Cambridge University Press, 2004.

[89] T. D. Wickens, *Elementary Signal Detection Theory.* Oxford University Press, USA, 2002.

[90] N. Douer and J. Meyer, Theoretical, Measured and Subjective Responsibility in Aided Decision Making, *Working Paper.* 2019 Available: https://arxiv.org/ftp/arxiv/papers/1904/1904.13086.pdf

[91] I. Pollack and A. B. Madans, "On the Performance of a Combination of Detectors," *Human Factors,* vol. 6, *(5),* pp. 523-531, 1964.